\documentclass[aps,twocolumn,nofootinbib,preprintnumbers,citeautoscript]{revtex4}
\usepackage{amsmath,amssymb,epsfig}
\usepackage{enumerate}

\def\be{\begin{equation}}
\def\ee{\end{equation}}

\def\be{\begin{equation}}
\def\ee{\end{equation}}

\def\beq{\begin{eqnarray}}\def\eeq{\end{eqnarray}}

\usepackage{graphicx}

\begin{document}
\title{Non-classical paths in interference experiments}

\author{{Rahul Sawant }$^{1}$,{ Joseph Samuel }$^{1}$,{ Aninda Sinha }$^{2}$,
{ Supurna Sinha }$^{1}$ and{ Urbasi Sinha }$^{1,3}$\\
\it $^1$Raman Research Institute, Sadashivanagar, Bangalore, India.\\
\it $^2$Centre for High Energy Physics, Indian Institute of Science,  Bangalore, India. \\
\it $^3$ Institute for Quantum Computing, 200 University Avenue West, Waterloo, Ontario, Canada.\\
\it{$^\ast$To whom correspondence should be addressed; E-mail:  usinha@rri.res.in.}}

\begin{abstract}

 In a double slit interference experiment, the wave function at the screen with both slits open is not exactly equal to the sum of the wave functions with the slits individually open one at a time. The three scenarios represent three different boundary conditions and as such, the superposition principle should not be applicable. However, most well known text books in quantum mechanics implicitly and/or explicitly use this assumption which is only approximately true. In our present study, we have used the Feynman path integral formalism to quantify contributions from non-classical paths in quantum interference experiments which provide a measurable deviation from a naive application of the superposition principle. A direct experimental demonstration for the existence of these non-classical paths is hard.   We find that contributions from such paths can be significant and we propose simple three-slit interference experiments to directly confirm their existence. 


\end{abstract}

\maketitle

\newpage



Quantum mechanics has been one of the most successful theories of the twentieth century, both in describing fundamental aspects of modern science as well as in pivotal applications. However, inspite of these obvious triumphs, there is universal agreement that there are aspects of the theory which are counter-intuitive and perhaps even paradoxical. Furthermore, understanding fundamental problems involving dark matter and dark energy \cite{dark1, dark2} in cosmology may need a consistent quantum theory of gravity. Unification of quantum mechanics and general relativity towards a unified theory of quantum gravity \cite{qg1, qg2} is the holy grail of modern theoretical physics. Such unification attempts involve modifications of either or both theories. However, all such attempts would rely very strongly on precise knowledge and understanding of the current versions of both theories. This makes precision tests of fundamental aspects of both quantum mechanics and general relativity very important to provide guiding beacons for theoretical development. 


The double slit experiment (figure 1) is one of the most beautiful experiments in physics. In addition to its pivotal role in optics, it is frequently used in classic textbooks on quantum mechanics \cite{book1,book2,book3} to illustrate basic principles. 
Consider a double slit experiment with incident particles (eg. photons, electrons). 
The wave function at the detector with slit A open is $\psi_A$. The wavefunction with the slit B open is $\psi_B$.
 What is the wavefunction with both slits open?  It is usually assumed to be $
 \psi_{AB} = \psi_A +\psi_B$ \cite{book1,book2,book3}. This is illustrated in figure 1. From the mathematical perspective of solving the Schr\"odinger equation, this assumption is definitely not
 true. The three cases described above correspond to three different boundary conditions \cite{yabuki, draedt} and as such the application of the superposition principle can at best be approximate. Recent numerical simulations of Maxwell's equations using Finite Difference Time Domain analysis have shown this to be true in the classical domain \cite{draedt}. How do we quantify this effect in quantum mechanics?
\begin{figure}
\centering
\includegraphics[scale=0.200]{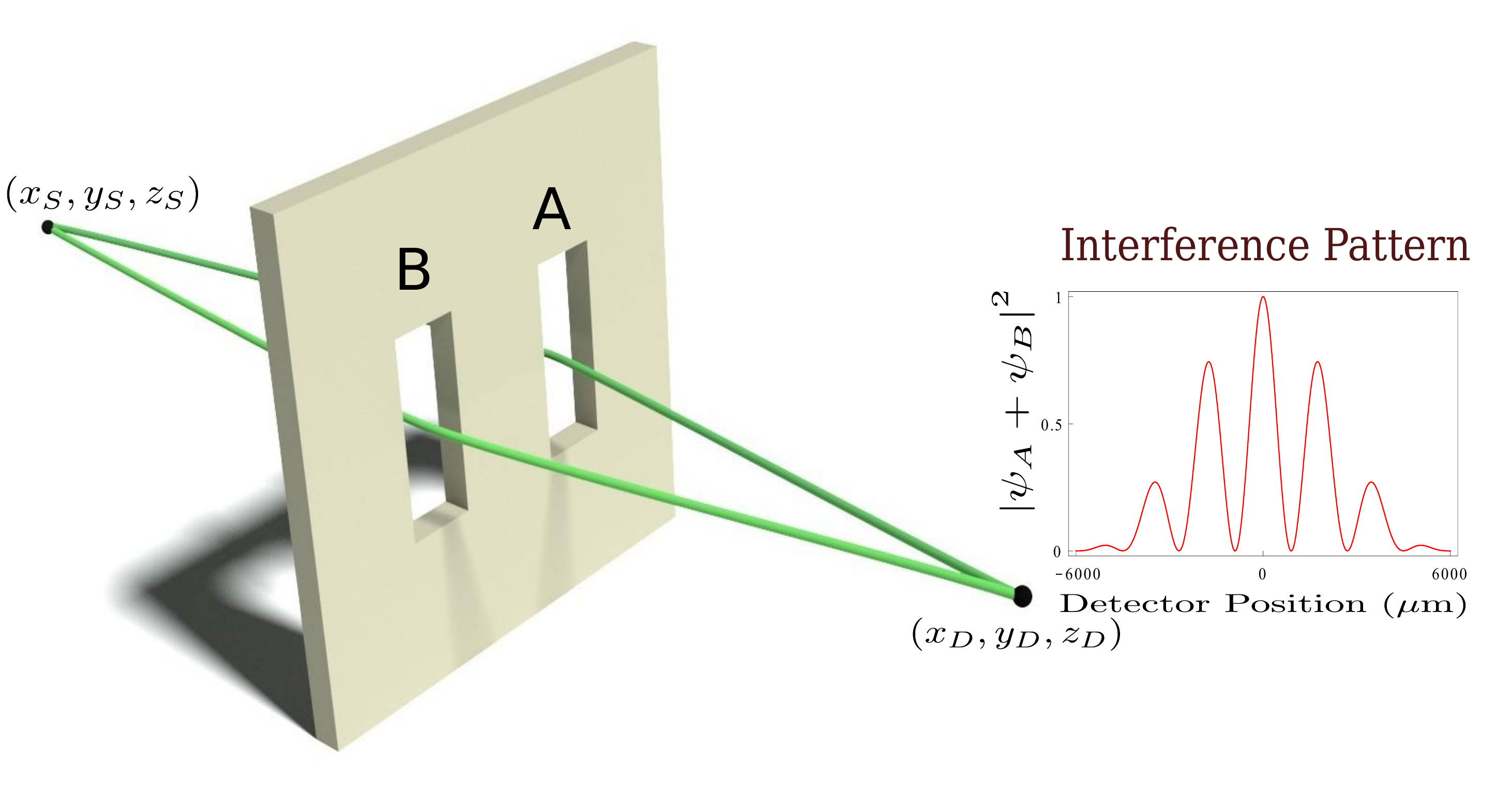}
\caption{The two slit experiment. The inset shows a typical interference pattern obtained by assuming $\psi_{AB} = \psi_A +\psi_B$.}
\label{}
\end{figure}

An intuitive and simple way of understanding this problem is to appeal to Feynman's path integral formalism \cite{feynman}. The path integral formalism involves an integration over all possible paths that can be taken by the particle through the two slits. This not only includes the nearly straight paths from the source to the detector through either slit (the classical paths) like the green paths in figure 2 but also includes paths of the type shown in purple in figure 2 (non-classical paths). These looped paths are expected to make a much smaller contribution to the total intensity at the detector screen as opposed to the contribution from the straight line paths. However, their contribution is finite. Formally, a classical path is one that extremizes the classical action. Any other path is a non-classical path.
This leads to a modification of the wave function at the screen which now becomes:
\begin{eqnarray}
     \psi_{AB} = \psi_A + \psi_B + \psi_L\,,
   \end{eqnarray}
where $\psi_L$ is the contribution due to the looped {\it i.e.,} non-classical paths. That $\psi_L$ is non-zero was first pointed out in \cite{yabuki} in the non-relativistic domain where certain unphysical approximations were made in computing $\psi_L$ and hence the results or the methods cannot be used in an experimental situation. Recently, \cite{draedt} has reiterated the point that $\psi_L$ can be non-zero without attempting to quantify it in quantum mechanics. 

\begin{figure}[h]
\centering
\includegraphics[scale=0.3]{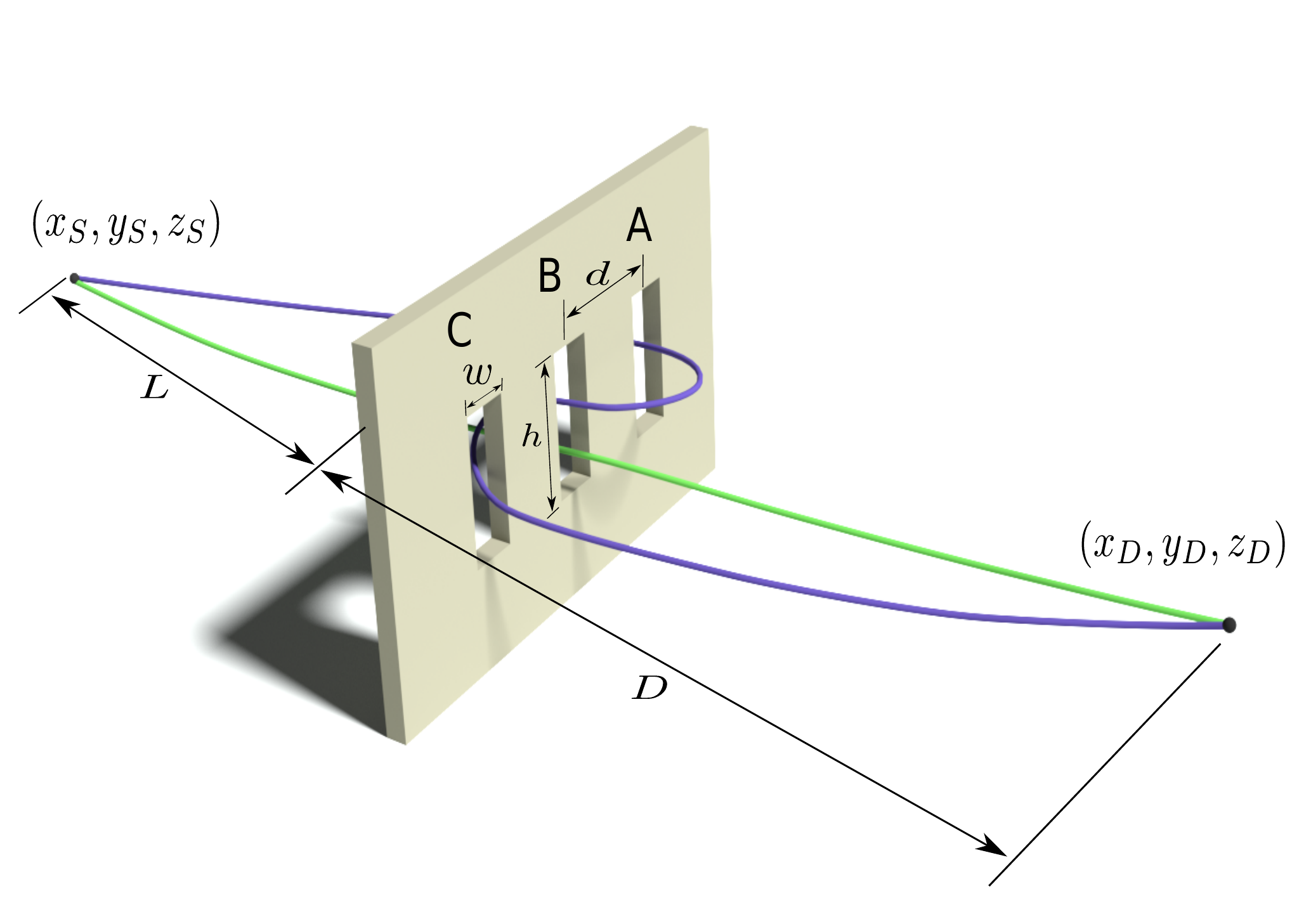}
\caption{Path integrals in a lab: The green line demonstrates a representative classical path. The purple line demonstrates a representative non-classical path. The various length parameters are marked; $d$ = the interslit distance, $w$ = the slit width, $h$ = the slit height, $L$ = the distance from the source to the slit plane and $D$ = the distance from the slit plane to the detector plane. }
\label{}
\end{figure}
In this paper, we will quantify the effect of such non-classical paths in interference experiments, thus quantifying the deviation from the common but incorrect application of the superposition principle in different possible experimental conditions. A well-known example of a direct experimental demonstration of such non-classical paths involves  the measurement of the Aharonov-Bohm phase \cite{ab}.  Berry's ``many-whirls" representation \cite{berry} provides insight into simple explanations of the Aharonov Bohm effect in terms of interference between whirling waves passing around the flux tube.  However, in most experimental attempts to measure the Aharonov Bohm phase, the detection relies on rather complicated experimental architecture and the results are also open to interpretational issues and further discussion \cite{ab1, ab2}. In this work, we propose simple triple slit based interference experiments \cite{usinha} which can be used as table top demonstrations of non-classical paths in the path integral formalism. Non-classical paths have been used to compute the semi-classical off-diagonal contributions to the two-point correlation function of a quantum system whose classical limit is chaotic \cite{richter}. The paths in this case are real. In the Feynman path integral approach, all possible paths going from the initial to final state need to be considered with an appropriate weight. In this sense all paths are real although in a physical quantity the contribution from certain paths may be suppressed.

The triple slit experiment provides a simple way to quantify the effects from non classical paths in terms of directly measurable quantities. The triple slit (path) set-up has been used as a test-bed for testing fundamental aspects of quantum mechanics over the last few years \cite{sorkin,usinha, sollner,park,niestegge,emerson}. Three-state systems are also fast becoming a popular choice for fundamental quantum mechanical tests \cite{zeilinger, piotr}. 
In order to analyse the effect of non-classical paths in interference experiments, we have considered the effect of such paths on an experimentally measurable quantity $\kappa$. $\kappa$ (defined below) has been measured in many experiments over the last few years in order to arrive at an experimental bound on possible higher order interference terms in quantum mechanics \cite{franson, hickmann} and in effect the Born rule for probabilities \cite{usinha, sollner,park}. Investigations of this quantity may also be relevant to theoretical attempts to derive the Born rule \cite{zurek}.
 If Born's postulate for a square law for probabilities is true and if $\psi_L=0$, then the quantity $\epsilon$ defined by
\begin{eqnarray}
\epsilon = p_{ABC}-(p_{AB}+p_{BC}+p_{CA})+(p_A+p_B+p_C) \,.
\label{eqn:sorkinnumber}
\end{eqnarray}
is  identically zero in quantum mechanics. 
Here $p_{ABC}$ is the probability at the detector when all three slits are open, $p_{AB}$ is the probability when slits A and B are open and so on.

In the experiments reported in the literature, the normalization factor has been chosen to be the sum of the three double slit interference terms called $\delta$ given by:
$
\delta = |I_{AB}|+|I_{BC}|+|I_{CA}|\,,
$
where $I_{AB} = p_{AB} - p_A - p_B$ and so on. This choice of normalization can sometimes lead to false peaks in the $\kappa$ as a function of detector position due to the denominator becoming very small at certain positions. We use a somewhat different normalization, $\delta=I_{max}$, where $I_{max}$ is the intensity at the central maximum of the triple slit interference pattern to avoid this problem.
Then the normalized quantity $\kappa$ is given by:
\begin{eqnarray}
\kappa = \frac{\epsilon}{\delta}\,.
\end{eqnarray}
In discussions which invoke the ``zeroness" of $\kappa$, it is implicitly assumed that only classical paths contribute to the interference. In his seminal work \cite{sorkin}, Sorkin had also assumed that the contribution from non-classical paths was negligible. 
Now, what is the effect of non-classical paths on $\kappa$? If one can derive a non-zero contribution to $\kappa$ by taking into account all possible paths in the Feynman path integral formalism, that would mean $ \psi_{AB} = \psi_A +\psi_B$ is not strictly true and experimentalists should not be led to conclude that a measurement of non-zero $\kappa$ would immediately indicate a falsification of the Born Rule for probabilities in quantum mechanics. A measured non-zero $\kappa$ could also be explained by taking into account the non-classical paths in the path integral. There is thus a theoretical estimate for a non-zero $\kappa$.  Of course, the immediate expectation would be a clear domination of the classical contribution and perhaps a very negligible contribution from the non-classical paths which would in turn imply that $ \psi_{AB} = \psi_A +\psi_B$ is true in all ``experimentally observable conditions." However, what we go on to discover is that this expectation is not always true. It is possible to have experimental parameter regimes in which $\kappa$ is measurably large. This in turn leads to a paradigm shift in such precision experiments. Observation of a non-zero $\kappa$ which is expected from the proposed correction to $ \psi_{AB} = \psi_A +\psi_B$ would in fact also serve as an experimental validation of the full scope of the Feynman path integral formalism. 

 As mentioned before, in calculating $\kappa$, one inherently assumes contributions only from the classical straight line paths as shown in green in figure 2. In this paper, we have estimated the contribution to $\kappa$ from non-classical paths, thus providing the first theoretical estimate for $\kappa$.

 For simplicity, we will use the free particle propagator in our calculations. For a particle in free space and away from the slits, this is a reasonable approximation. We account for the slits by simply removing from the integral all paths that pass through the opaque metal. An estimate for the error due to this assumption has been worked out in \cite{support}. The normalized energy space propagator $K$ \cite{support} for a free particle with wave number $k$ from a position $\vec{r'}$ to $\vec{r}$ is given by 
\begin{eqnarray}
K(\vec{r},\vec{r}') =\frac{k}{2\pi i}  \frac{1}{|\vec{r}-\vec{r}'|}e^{ik|\vec{r}-\vec{r}'|} \,.
\label{eqn:propagator}
\end{eqnarray}
 Although in this paper, we will be mainly focusing on analyzing optics based experiments using photons, this propagator equation can be used both for the electron and the photon as argued in \cite{support}. We should point out that there are corrections to the propagator due to closed loops in momentum space from quantum field theory considerations. We have explicitly estimated that the effects of such corrections will be negligibly small \cite{comment3}. 

Consider the triple slit configuration shown in figure 2. 
According to the path integral prescription, all paths that go from the
source to the detector should contribute in the analysis. In the quantity
of interest, $\kappa$, some important simplifications occur. Only those
non-classical paths that involve propagation between at least two slits would contribute
to the leading non-zero value. This is because any non-classical path that goes
through {\it only} the $i$'th slit can be taken into account in the wave-function $\psi_i$ at the detector and hence would cancel
out in $\kappa$ as can be easily checked.
In light of the above, the entire set of paths from the source to the detector through the slits can be divided into two classes:

\begin{enumerate}

\item Paths which cross the slit plane exactly once pertaining to a probability amplitude $K_{c}$; a representative path is shown by the green line and 

\item Paths which cross the slit plane more than once at two or more slits pertaining to a probability amplitude $K_{nc}$ \cite{support} as for instance, represented by the purple line.

\end{enumerate}

\begin{eqnarray}
\therefore K= K_{c} + K_{nc}\,.
\end{eqnarray}

We wish to estimate $K_{nc}$ relative to $K_{c}$. An example of a representative  $K_{c}$ in our problem is the probability amplitude to go from the source $(-L,0,0)$ to the detector $(D,y_D,0)$ through slit $A$ which we call $K^A(S,D,k)$. This uses the general scheme that a path in Feyman's path integral formalism can be broken into many sub-paths and the propagator is the product of the individual propagators \cite{support}. For instance, 
\begin{eqnarray}
K^A_c&=& - (\frac{k}{2\pi})^2\int_{d-\frac{w}{2}}^{d+\frac{w}{2}}\int_{-h}^{h} \ dy \ dz \ \frac{e^{ik(l_1+l_2)}}{l_1l_2} \,, 
\end{eqnarray}
where $d$ is the interslit distance, $w$ is the slit width, $h$ is the slit height, $l_1^2 = y^2+L^2+z^2$ and $l_2^2 = (y_D-y)^2+D^2+z^2$ as shown in figure 2. For the source and the detector far apart from one another, {\it i.e.,} in the Fraunhofer regime, $D\gg d$ $\&$ $L\gg d$ in the region of integration, therefore, $ l_1 \approx L + \frac{y^2+z^2}{2L} $. Similarly
$
l_2^2 = (y_D-y)^2+D^2+z^2$ giving $l_2 \approx D + \frac{(y_D-y)^2+z^2}{2D}$. Thus we have
\begin{eqnarray}
K^A_c &=& -\gamma (\frac{k}{2\pi})^2 \int_{d-\frac{w}{2}}^{d+\frac{w}{2}}\int_{-h}^{h} \ dy \ dz \ e^{ik[ \frac{y^2+z^2}{2L} + \frac{(y_D-y)^2+z^2}{2D}]}\,.\nonumber\\
\end{eqnarray}
Here $\gamma=\frac{1}{LD} e^{ik(L+D)}$.
These are Fresnel integrals and have been evaluated using Mathematica.

Let us now proceed to the probability amplitude for multiple slit crossings {\it i.e.,} $K_{nc}$. An example of a representative  $K_{nc}$ in our problem is the probability amplitude to go from the source $(-L,0,0)$ to the detector $(D,y_D,0)$ following the kind of path shown in figure 2. In this case, the particle goes from the source to the first slit and then loops around the second and third slits before proceeding to the detector. We represent this by $K_{nc}^A(S,D,k)$. This is approximated by \cite{support}:
\begin{equation}
\begin{split}
K_{nc}^A  = i (\frac{k}{2\pi})^3 \int\ dy_1 \ dy_2 \ dz_1 \ dz_2 \ \frac{e^{ik(l_1+l_2+l_3)}}{l_1l_2l_3}  \,.
\end{split}
\end{equation}
Here the $y_1$ integral runs over slit $A$ and $y_2$ integral runs over slits $B$ and $C$ and where
$l_1^2 = (y_1-y_S)^2+L^2+z_1^2,~~  l_2^2 = (y_2-y_1)^2+(z_2-z_1)^2\,,{\rm and~}  l_3^2 = (y_D-y_2)^2+D^2+z_2^2.$
Making approximations appropriate to the Fraunhofer regime, using stationary phase approximation \cite{stationary}  for the oscillatory integrals the integral becomes:
\begin{widetext}
\begin{eqnarray}
K_{nc}^A &=& \gamma i^{3/2}(\frac{k}{2\pi})^{5/2} \int\ dy_1 \ dy_2  \ dz_1 \, |y_2-y_1|^{-1/2}e^{ik\left[\frac{y_1^2+z_1^2}{2L}+(y_2-y_1)+\frac{(y_D-y_2)^2+z_1^2}{2D}\right]}\,.
\end{eqnarray}
\end{widetext}
An important simplification occurs at this stage: the $z$ integral in $K_{nc}^A$ is same as in the integral for $K_{c}^A$. Since we are just concerned with ratios, the contributions from the $z$ integrals cancel out. 
%

In terms of $K_{c}$ and $K_{nc}$, the propagator to go from the source to the detector when all three slits are open is given by:
\begin{eqnarray}
K^{ABC} = K^A_c + K^B_c + K^C_c + K_{nc}^{ABC}\,,
\end{eqnarray}
where $K_{nc}^{ABC}$ include non-classical terms arising when all slits are open.
Similarly:
\begin{eqnarray}
K^{AB} = K^A_c + K^B_c +K_{nc}^{AB}\,.
\end{eqnarray}
$K_{nc}^{AB}$ are non-classical terms involving only $A$ and $B$.
Similarly for $AC$ and $BC$.
Thus, in terms of propagators,
\begin{eqnarray}
\epsilon &=& 	|K^{ABC}|^2-|K^{AB}|^2-|K^{AC}|^2-|K^{BC}|^2 \nonumber \\ &&~~~~~~~~~~~~+ |K^A|^2+|K^B|^2+|K^C|^2
\label{eqn:kappa} \,,
\end{eqnarray}
%
\begin{figure}[h]
\centering
\includegraphics[scale=0.35]{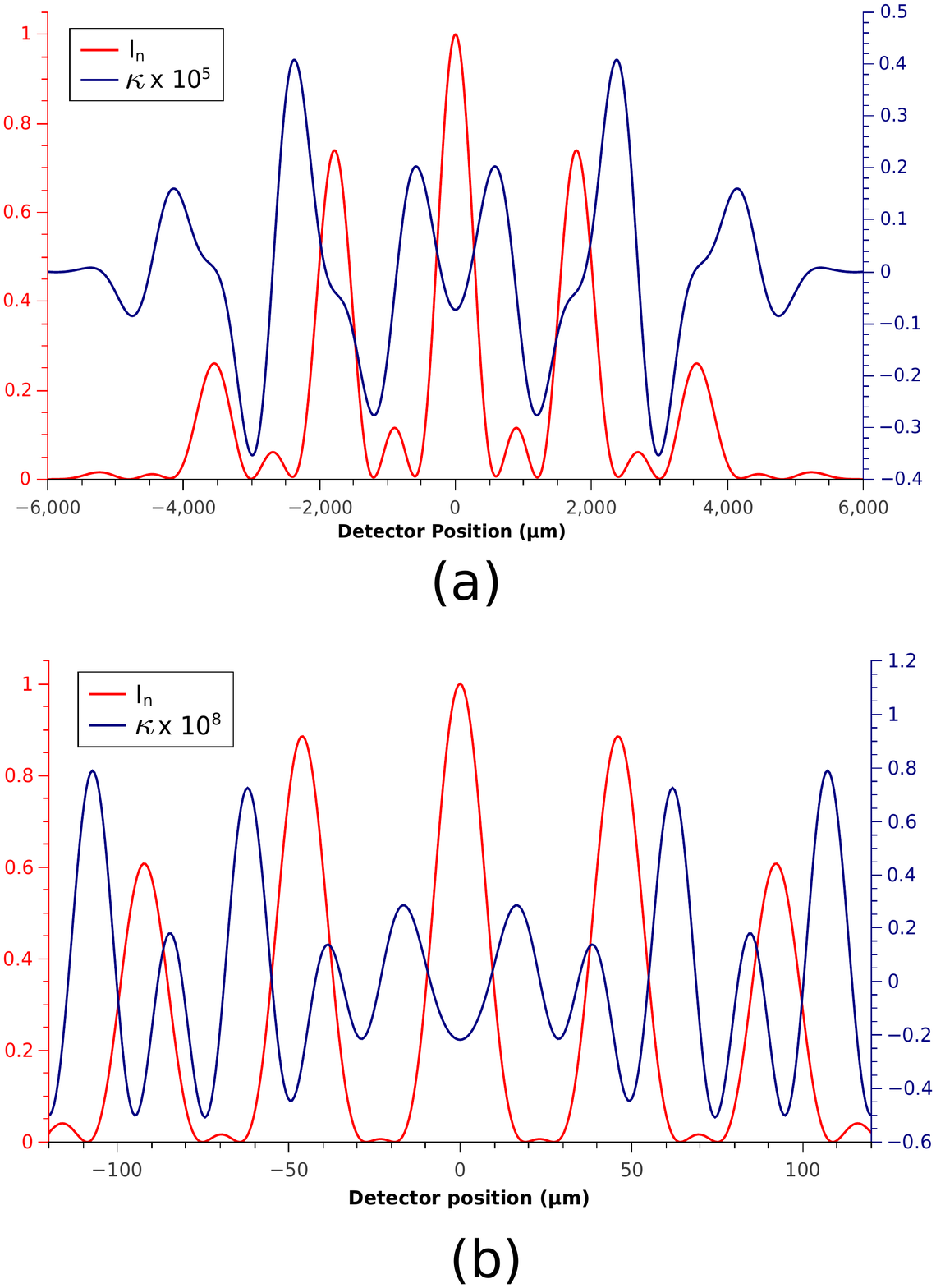}
\caption{Normalised values of $\kappa$ as a function of detector position. Here $I_n = |K^{ABC}(y)|^2/|K^{ABC}(0)|^2$.  a) This is for incident photons, slit width = 30$\mu$m, interslit distance = 100$\mu$m, distance between source and slits and slits and detector = 18cm and incident wavelength = 810nm \cite{usinha}. b) This is for incident electrons,  slit width = 62nm, interslit distance = 272nm, distance between source and slits = 30.5cm and slits and detector = 24cm and deBroglie wavelength = 50pm \cite{electron}.}
\label{}
\end{figure}
%
and the normalization $\delta$ is given by $\delta = |K^{ABC}(0)|^2$, where $|K^{ABC}(0)|^2$ is the value of $|K^{ABC}|^2$ at the central maximum.
By numerical integration, we find $\kappa$ at the central maximum of the triple slit interference pattern to be of the order of 
$10^{-6}$ for the parameters used in the triple slit experiment reported in reference \cite{usinha}. What would have been expected to be zero considering only straight line paths now turns out to be measurably non-zero having taken the non-classical ones into account \cite{comment4}.
In figure 3, we show $\kappa$ as a function of detector position. We also show a plot of the triple slit interference pattern as a function of detector position which gives a clearer understanding of the modulation in the plot for $\kappa$.


The experiment reported in reference \cite{usinha} was not sensitive to a theoretically expected non-zeroness in $\kappa$ due to systematic errors. However, in the absence of such systematic errors, it is definitely possible to use a similar set-up to measure a non-zero $\kappa$. Simulation results indicate that the set-up could have measured a much lower value of $\kappa$ but the presence of the systematic error due to one misaligned opening in the blocking mask set the limitation of the experiment making it possible to only measure a value of $\kappa$ upto $10^{-2}$. There is no reason why this systematic error cannot be removed in a future version of the experiment thus making it a perfect table-top experiment to test for the presence of non-classical paths in interference experiments.  However, experiments of the kind reported in \cite{sollner} are not as ideally suited for this purpose. This is because, in our analysis, we have worked in the thin-slit approximation. The effective ``slit-thickness'' in a diffraction grating based interferometer set-up would be quite big and hence the resulting $\kappa$ would certainly be smaller. 

What we go on to also find in our current analysis is that $\kappa$ is very strongly dependent on certain experimental parameters and one can definitely find a parameter regime where $\kappa$ would be even bigger, hence easier to observe. We find that keeping all other experimental parameters fixed, $\kappa$ increases with an increase in wavelength. Thus, for instance, for an incident beam of wavelength 4cm (microwave regime) and slit width of 120cm and interslit distance of 400cm, a theoretical estimate for $\kappa$ would be $10^{-3}$. This is an experiment which can be performed for instance in a radio astronomy lab.

Experiments of this kind where the value of $\kappa$ due to non-classical paths can be estimated would definitely be of great interest as they would serve as a simple experimental demonstration of how the basic assumption that a composite wavefunction is just the sum of component wavefunctions is not always true. In a sense they would also serve as a direct table-top demonstration of the complete scope of the Feynman path integral formalism where not only the straight line paths are important but also the looped paths can make a sizeable contribution depending on one's choice of experiment. The effects due to such non-classical paths may also be used to model possible decoherence mechanisms in interferometer based quantum computing applications. 


\section*{Acknowledgments} We thank Aveek Bid, Dwarkanath K.S., Subroto Mukerjee, Robert Myers, Rajaram Nityananda, Barry Sanders,  Rafael Sorkin, Radhika Vatsan and Gregor Weihs for useful discussions. We thank Raymond Laflamme and Anthony Leggett for reading through the draft and helpful comments and discussions. AS acknowledges partial support from a Ramanujan fellowship, Govt. of India.


\renewcommand{\thesection}{S.\arabic{section}}
\renewcommand{\thesubsection}{\thesection.\arabic{subsection}}
 
%
\makeatletter 
\def\tagform@#1{\maketag@@@{(S\ignorespaces#1\unskip\@@italiccorr)}}
\makeatother
 
\makeatletter
\makeatletter \renewcommand{\fnum@figure}
{\figurename~S\thefigure}
\makeatother





\newpage

\begin{center}
{\bf Supplementary material }
\end{center}

\begingroup
\let\newpage\relax
\maketitle
\endgroup
\section{Assumptions}

\subsubsection*{Stationary Experiments}
We suppose that we have a monochromatic source of light
(monoenergetic source of electrons) and that the detectors integrate over 
the
duration $T$ of the experiment.  Assuming that $T$ is much longer
than any other time scale in the problem, like the travel
time across the apparatus, we can use a steady state description.
Both electrons and light are then described (in a scalar approximation)
by the Helmholtz equation 
\begin{eqnarray}
\left(\nabla^2+ k^2 \right)\psi_k(\vec{r}) = 0,  
\end{eqnarray} 
which is satisfied away from the sources
and detectors. $\psi_k$ is a scalar field representing the wave function
of the electron or  a component of the electromagnetic vector potential.
For light $k=\omega/c$ and for electrons $k=\sqrt{2m_e E}$ ( setting $\hbar=1$ ).
Both electron and photon diffraction can be treated on the same footing
in the time independent case. Below we will drop the superscript $k$
on physical quantities to avoid cluttering the formulae with it.
We will also suppose throughout that $\lambda$ is much smaller than
any other length scale in the problem, the sizes and separations of the slits and the
distance to the source and the detector.

\subsubsection*{Free Propagation and Huygens principle}
To describe free propagation, we use the Kernel (see section 59  of \cite{landau})
\begin{equation}
K({\vec r}_1,{\vec r}_2)=\frac{k}{2\pi i} 
\frac{e^{ik|{\vec r}_1-{\vec r}_2|}}{|{\vec r}_1-{\vec r}_2|},
\label{kern}
\end{equation}
which 
\begin{enumerate}[(i)]
\item satisfies the Helmholtz equation\cite{bornwolf} away from
$\vec{r_1}=\vec{r_2}$ and
\item satisfies the Fresnel-Huygens principle:
\begin{equation}
K({\vec r}_1,{\vec r}_3)=\int d{\vec r}_2 
K({\vec r}_1,{\vec r}_2)K({\vec r}_2,{\vec r}_3)
\label{fold}
\end{equation}
where $\vec{r}_2$ is integrated over any plane between ${\vec r}_1$ 
and ${\vec r}_3$
\cite{landau,bornwolf} and perpendicular to ${\vec r}_1-{\vec r}_3$. 
\end{enumerate}
\subsubsection*{Use of time independent Feynman path integral}
By repeated application of equation S\ref{fold}, we can express
the propagator for free space in the form
\begin{equation}
K({\vec r}_1,{\vec r}_2)=\int {\cal D} [{\vec x}(s)] \exp{[ik\int ds]},
\label{kpath}
\end{equation}
where $s$ is the contour length along the path ${\vec x}(s)$ and the sum is
over all paths connecting $\vec{r_1}$ with $\vec{r_2}$. 
In the classical limit of $k\rightarrow \infty$, paths near the straight 
line path joining $\vec{r_1}$ to $\vec{r_2}$ contribute by stationary 
phase. We refer to these as ``classical paths'' in the text. All of these
would contribute ``in phase''. Paths away from the classical path are
expected to contribute with rapidly oscillating phase.
In describing  diffraction by a system of slits, we would have 
to sum over all paths
connecting the source to the detector. This would include paths of
the kind shown in purple in Fig. 2, which are far from classical. 
We would expect (in the limit of small $\lambda$)
that the contributions of such paths are negligible because of 
rapid oscillations of the phase.  We would like to know just how ``small" these
contributions are.

\section{Contribution of classical and non-classical paths to the kernel}

In Fresnel's theory of diffraction by a slit \cite{kumar} we use equation S\ref{fold} to 
insert a single intermediate state on the slit plane and find 
the amplitude $K_1$ to reach $D$ from $S$:
\begin{equation}
K_1^\Omega({\vec r}_S,{\vec r}_D)=\int_\Omega d{\vec r} 
K({\vec r}_S,{\vec r})K({\vec r},{\vec r}_D),
\label{ksd}
\end{equation}
where ${\vec r}$ is an intermediate point on the slit plane (taken to be the 
$(y,z)$ plane). The range of integration in equation S\ref{ksd} is 
over the {\it two dimensional} region $\Omega$ in the slit plane, 
where $\Omega$ is the union of the open slits. 
This gives a remarkably accurate account of the phenomenon of diffraction.  
In Fresnel's theory we find because there is a single 
integration over $\Omega$, that the outcomes of the three possible
experiments with two slits (A, B and AB open) are related by \cite{book1,book2,book3}
\begin{equation}
K_1^{AB}=K_1^A+K_1^B.
\label{ksum}
\end{equation}

Going beyond Fresnel's theory, 
we insert two intermediate points on the slit plane and 
integrate twice\cite{footnote, frmp} over $\Omega$, the open parts of the slit plane. 
\begin{equation}
K_2^\Omega({\vec r}_S,{\vec r}_D)=\int_\Omega d{\vec r}_1\int_\Omega d{\vec r}_2 
\,\,K({\vec r}_S,{\vec r}_1)K({\vec r}_1,{\vec r}_2)
K({\vec r}_2,{\vec r}_D).
\label{ktwo}
\end{equation}
A typical path in this integration has two 
``kinks'' (at ${\vec r}_1$ and ${\vec r}_2$) and thus the
integral is highly oscillatory. These integrals in $K_2^\Omega$ can 
be computed numerically and seen to be much smaller than $K_1^\Omega$. 
In this order of approximation the amplitude for detection 
at $D$ is given by 
\begin{equation}
K^\Omega({\vec r}_S,{\vec r}_D)=K_1^\Omega({\vec r}_S,{\vec r}_D)+K_2^\Omega({\vec r}_S,{\vec r}_D) + \ . \ . \ . \ . \ . \ .\ .\ .\ , 
\label{total}
\end{equation}
where $K_2$ is the leading order contribution to the amplitude due to non-classical paths $K_{nc}$.  As mentioned before $K_2^\Omega$ is expected to be much smaller than
$K_1^\Omega$. Most importantly, because of the two integrations
over $\Omega$ in equation S\ref{ktwo}, $K_2^\Omega$ results  in violations of equation {\ref ksum}.

Sorkin suggested that the Born rule for probabilities in quantum mechanics can be tested by performing a three slit experiment\cite{sorkin}. 
By keeping each slit either open or closed, we can perform seven distinct
non trivial experiments. Theoretical predictions for the outcomes of
these seven experiments are given by choosing $\Omega$ to be one of
the seven domains $\{A,B,C, A\cup B,B\cup C,C\cup A, A\cup B \cup C\}$.
Following Sorkin, we define the quantity
\begin{eqnarray}
\epsilon&=&|K^{ABC}|^2-(|K^{AB}|^2+|K^{BC}|^2+|K^{CA}|^2)\nonumber\\
&& + \ |K^{A}|^2+|K^B|^2+|K^C|^2.
\label{epsilon}
\end{eqnarray}
A straightforward calculation shows that 
after cancellations and to linear order in $K_2$, 
\begin{eqnarray}
\epsilon&=&2 {\rm{Re}}\{\overline{K_1^C}(K_2^{AB}+K_2^{BA})+
\overline{K_1^{A}}(K_2^{BC}+K_2^{CB})\nonumber \\
&&+ \ \overline{K_1^B}(K_2^{AC}+K_2^{CA})\}.
\label{final}
\end{eqnarray}
This final expression shows clearly that it is the $K_2$ terms that
violate $ \psi_{AB} = \psi_A + \psi_B$ and make $\epsilon$ non zero. In this equation $K_1$ would also receive contributions from non-classical paths, for example one which crosses the slit plane exactly once, but has a kink in it. However, it is evident that these would contribute at a sub-leading order to a non-zero $\epsilon$. $\kappa=\epsilon / \delta$ has been computed numerically and the resulting graphs are
shown in Fig.3.  

There are several subtleties associated with Huygens principle,
which do not however affect the order of magnitude we get for $\kappa$.
Huygens initially gave a construction for evolving the wavefront
using secondary wavelets. It was Fresnel \cite{bornwolf} who applied
Huygen's construction to understand diffraction effects. However, Fresnel
had to introduce {\it ad hoc} some ``inclination factors''. Subsequently
Kirchoff derived Fresnel's theory from Maxwell's equations
using an integral equation derived from Helmholtz's equation (see equation 17 on 
page 422 of \cite{bornwolf}). He was also
able to derive (and correct) Fresnel's ``inclination factors''. In equation S\ref{ktwo}
these factors result in an additional factor of $1/4$ (because of two right angle kinks)
and this leads to a factor of $1/4$ multiplying $\epsilon$.

Classically the path taken by a particle is a path of least action. 
But quantum mechanics tells us that each physically possible path 
has a probability amplitude associated to it. The final probability 
amplitude is summation of probability amplitudes from all 
paths \cite{feynman}.\\   
In solving the problem of scattering due to the presence of 
slits using Feynman path integral (equation S\ref{kpath}) we simply excise all those paths
which go through the solid portion of the slits. 
In Fresnel's theory we suppose that the 
amplitude at slit $A$ is the same as it would be in free space. 
In the next order, we allow for the fact that the amplitude
at $A$ could also be influenced by waves arriving through slit $B$
(if $B$ is open). This is why the non-classical effects violate
the simple minded application of the superposition principle.


\section{Details of calculations}
The classical amplitude is given by,
\begin{eqnarray}
K^A_c &=& -\gamma (\frac{k}{2\pi})^2 \int_{d-\frac{w}{2}}^{d+\frac{w}{2}}\int_{-h}^{h} \ dy \ dz \ e^{ik[ \frac{(y)^2+z^2}{2L} + \frac{(y_D-y)^2+z^2}{2D}]}\,.\nonumber\\
\end{eqnarray}

Here the integrals for $z$ and $y$ variables are independent. The $z$ integral evaluates to a complex number,
\begin{eqnarray}
C_z &=& \int_{-h}^{h}  \ dz \ e^{ik[ \frac{z^2}{2L} + \frac{z^2}{2D}]}\,.\nonumber\\
\end{eqnarray}
This is same for all the slits $A,B$ and $C$. 

The non-classical amplitude within the Fraunhofer approximation is, 
\begin{widetext}
\begin{eqnarray}
\label{befst}
K_{nc}^A &=& \gamma i(\frac{k}{2\pi})^{3} \int\ dy_1 \ dy_2  \ dz_1 \ dz_2\  [(y_2-y_1)^2+(z_2-z_1)^2]^{-1/2}  e^{ik\left[\frac{y_1^2+z_1^2}{2L}+\sqrt{(y_2-y_1)^2+(z_2-z_1)^2} +\frac{(y_D-y_2)^2+z_2^2}{2D}\right]} \ \ \ \ \ \ \ \ \ \
\end{eqnarray}
\end{widetext}

For the $z_2$ integral using the stationary phase approximation (S\ref{stationary}) with $z_2 = z_1$ as stationary point and $g''(z_2) = \frac{1}{|y_2-y_1|}$ the non-classical amplitude reduces to,
\begin{widetext}
\begin{eqnarray}
K_{nc}^A &=& \gamma i^{3/2}(\frac{k}{2\pi})^{5/2} \int\ dy_1 \ dy_2  \ dz_1 \, |y_2-y_1|^{-1/2}e^{ik\left[\frac{y_1^2+z_1^2}{2L}+|y_2-y_1|+\frac{(y_D-y_2)^2+z_1^2}{2D}\right]}\,.
\end{eqnarray}
\end{widetext} 

Again the integrals for $z$ and $y$ variables are independent and the $z_1$ integral is equal to $C_z$. As the numerator and denominator in $\kappa$ are both linear combination of multiplications of type $K_{nc}^A\overline{K^A_c} $, the factor $|C_z|^2$ cancels out.  
\section{Detailed discussion on Errors}

\subsubsection*{Transmission through metal} 
We assume that the penetration of light through the opaque 
metal is zero. The transmission amplitude can be found heuristically.
If $\alpha$ is the attenuation constant and 
if $\psi_i$ is the incident wave amplitude the transmission 
amplitude is given by, $\psi_t = e^{-2 \pi  \alpha  \zeta}\psi_i$. $\zeta$ is the thickness of the layer in units of the wavelength. This quantifies the approximation that there is no path passing through the solid metal. 
Here $\psi_i \approx K^A$, therefore the transition amplitude is, $e^{-2\pi\alpha \zeta}K^A $. $\alpha = 2.61$ as refractive index of steel is $2.29 + 2.61i$.  $\zeta = 1 \mu m$, therefore the error is $K^A\times 10^{-8} $

\subsubsection*{Error due to stationary phase method} We assumed that the propagator from the source to the slit and the slit to the detector is a free particle propagator. To quantify this approximation we integrate over an intermediate plane ($x = L/2$) similar to equation S\ref{fold}. The integral is of the form,
\begin{eqnarray}
I = \int_{-\infty}^{\infty} f(y) e^{ikg(y)} \ dy .
\end{eqnarray}
Here,  $g(y)$ is the total distance from the source to the slit. $y$ is a variable .The stationary point for the above integral is a point lying on the straight line joining the source and the detector. 

A Taylor series expansion around a stationary point $y_0$, retaining the first two non-zero terms in series gives, 
\begin{equation}
I = \int f(y_0)e^{ik\left[g(y_0)+\frac{g_2(y_0)(y-y_0)^2}{2!}+\frac{g_4(y_0)(y-y_0)^4}{4!}\right]} dy. 
\end{equation}

 $g_4(y_0)$ denotes the fourth derivative of $g(y)$ at point $y_0$.
Performing the integral explicitly we get,
\begin{equation}
\label{stationary}
I = f(y_0)e^{ikg(y_0)}\sqrt{\frac{\pi}{kg''(y_0)}}e^{i\pi/4} \left[1+O\left(\frac{g_4(y_0)}{g_2(y_0)^{2}k}\right)\right].
\end{equation}

For our purpose , $g(y) = \sqrt{(L-L/2)^2+y^2}+\sqrt{(L/2)^2+(y-d)^2}$, $y_0=d/2$, $g_2(y_0) = \frac{4L^2}{(d^2+L^2)^{3/2}} \approx 4/L$ as $L \gg d$  and  $|g_4(y_0)| = \frac{48L^2(L^2-4d^2)}{(d^2+L^2)^{7/2}}\approx 48/L^3 $.\\
Therefore,
\begin{eqnarray}
\frac{|g_4(y_0)|}{g_2(y_0)^{2}k} = \frac{3}{Lk}. 
\end{eqnarray}
For $k= \frac{2 \pi}{\lambda}$ and $L = 10^5 \lambda$ , the error is $K_A \times 10^{-6}$

\subsubsection*{Error due to Fraunhofer approximation } In the Fraunhofer limit $L \gg d $, the errors due to this approximation are of the order $K_A \times d/L \approx K_A \times 10^{-4} $. \\

These errors result in error in calculating $\kappa$. The final leading order error in $\kappa$ is $\kappa \times 10^{-4}$. 

\vskip 2cm

\begin{center}
\line(1,0){250}
\end{center}
 
\renewcommand{\figurename}{Figure}


\begin{thebibliography}{}
\bibitem{comment3} Quantum field theory effects lead to a well known nonlinear correction to Maxwell equations arising from the Euler-Heisenberg lagrangian \cite{euler}. The contribution to $\kappa$ from this correction is found to be of the order $10^{-25}I$ where $I$ is the intensity measured in SI units. For instance, for a typical Titanium-Sapphire laser with power output 10 Watts and beam diameter 20$\mu$m, the correction to $\kappa$ is around 10 orders of magnitude smaller than what is reported here.
\bibitem{comment4} We have verified that in the absence of non-classical terms, the parameter $\kappa$ evaluates to zero within numerical accuracy.
\end{thebibliography}
\begin{thebibliography}{}
\bibitem{footnote}
Strictly Speaking , in equation S\ref{ktwo} one should excise the region where $|\vec{r_1}-\vec{r_2}| \sim \lambda$ (the diagonal of the $\Omega \times \Omega$ integration). This integral is already accounted for in $K_1^{\Omega}$. However this does not affect any of the terms in the $\epsilon$ expression (S\ref{final}).
\end{thebibliography}
\end{document}